\documentclass[pra,twocolumn]{revtex4-1}
\usepackage{hyperref} 

\usepackage{graphicx} 
\usepackage{subfig}
\usepackage{color}
\usepackage{filecontents}
\usepackage{soul}
\usepackage{mathrsfs}
\usepackage{tikz,bm} 
\usepackage{circuitikz}
\usepackage{tikz}
\usepackage{physics}
\usepackage{mathtools}
\usepackage[T1]{fontenc}
\usepackage[outline]{contour}
\usepackage{relsize}

\usetikzlibrary{shapes.geometric}
\usetikzlibrary{decorations.pathmorphing}
\usetikzlibrary{shapes.arrows, fadings}
\usetikzlibrary{arrows,snakes,shapes}

\definecolor{mycolor}{rgb}{1,0.2,0.3}
\definecolor{brightgreen}{rgb}{1.0, 1.0, 1.0}
\definecolor{britishracinggreen}{rgb}{0.0, 0.26, 0.15}
\definecolor{cadmiumgreen}{rgb}{0.0, 0.42, 0.24}
\definecolor{ceruleanblue}{rgb}{0.16, 0.32, 0.75}
\definecolor{darkelectricblue}{rgb}{0.33, 0.41, 0.47}
\definecolor{darkpowderblue}{rgb}{0.0, 0.2, 0.6}
\definecolor{dt}{rgb}{1.0, 0.66, 0.07} %darktangerine
\definecolor{emerald}{rgb}{0.31, 0.78, 0.47}
\definecolor{palatinatepurple}{rgb}{0.41, 0.16, 0.38}
\definecolor{pastelviolet}{rgb}{0.8, 0.6, 0.79}
\definecolor{br}{rgb}{0.5, 0.05, 0.01}
\definecolor{chosen_color}{RGB}{3, 207, 252}

\pgfdeclareverticalshading{pgf@lib@fade@north}{100bp}
{color(0bp)=(pgftransparent!0);
 color(30bp)=(pgftransparent!0);
 color(50bp)=(pgftransparent!80); 
 color(65bp)=(pgftransparent!100);
 color(100bp)=(pgftransparent!100)}%
\pgfdeclarefading{exponentialtop}{%
\pgfuseshading{pgf@lib@fade@north}%
}

\pgfdeclareverticalshading{pgf@lib@fade@north}{100bp}
{color(0bp)=(pgftransparent!100);
 color(35bp)=(pgftransparent!100);
 color(50bp)=(pgftransparent!80); 
 color(70bp)=(pgftransparent!00);
 color(100bp)=(pgftransparent!00)}%
\pgfdeclarefading{exponentialbottom}{%
  \pgfuseshading{pgf@lib@fade@north}%
}

\newcommand{\be}{\begin{equation}}
\newcommand{\ee}{\end{equation}}
\newcommand{\bea}{\begin{eqnarray}}
\newcommand{\eea}{\end{eqnarray}}
\newcommand*{\myeqref}[2][Eq.~]{%
\hyperref[{#2}]{#1(\ref*{#2})}%
}
\def\equationautorefname#1#2\null{%
Eq.#1(#2\null)%
}

\usepackage{dcolumn}
\usepackage{bm}
\usepackage{amssymb,amsmath}
\usepackage{color}
\usepackage{float}
\usepackage{tikz}
\usetikzlibrary{arrows}

\definecolor{DarkGreen}{rgb}{0,0.6,0.2}

\usepackage{mathrsfs}
\begin{document}
\title{Hong-Ou-Mandel interferometry with cavity QED-based single-photon sources: A Quantum Jump Analysis}
\author{Lexi Dudones} 
\author{Caden McCollum}  
\author{Imran M. Mirza}
\email{mirzaim@miamioh.edu}
\affiliation{Macklin Quantum Information Sciences,Department of Physics, Miami University, Oxford, OH 45056, USA}

%%===================================================%%
%%                   Article Abstract                %%
%%===================================================%%
\begin{abstract}
We present a quantum jump/trajectory analysis of the two-photon interference phenomenon in the context of the Hong-Ou-Mandel effect (HOME). In particular, we consider the standard setup of HOME, which consists of two-photon sources firing single photons from the opposite sides of a 50/50 beam splitter and two perfect detectors placed at the output ports to record photodetection events. For single-photon sources, we consider two special cases: (1) two excited two-level atoms and (2) two atom-cavity setups with initially present single photons in both cavities. For both cases, we report analytic results as well as quantum jump-based Monte Carlo simulations to demonstrate the signatures of these single-photon sources on the HOME under different working conditions (for example, strong- and weak-coupling regimes of cavity quantum electrodynamics). Our results may have interesting applications in linear optics quantum computing as well as in protocols that test the indistinguishability of single-photon sources. 
\end{abstract}
%%%%%%%%%%%%%%%%%%%%%%%%%%%%%%%%%%%%%%%%%%%%%%%%%%%%%%%

\maketitle

%%===================================================%%
%%         Sec. I: Article Introduction              %%
%%===================================================%%
\section{\label{sec:I} Introduction}  
The Hong-Ou-Mandel effect (or HOME) is a pure two-photon interference effect, which was originally reported by Hong, Ou, and Mandel in their seminal paper in 1987 \cite{hong1987measurement}. In their experiment, they found that only two of the four classically predicted outcomes are observed when two indistinguishable photons are fired at a 50/50 beam splitter within a very small time interval (on the order of picoseconds) \cite{bouchard2020two, branczyk2017hong}. Since then, HOME has been extensively studied in a wide range of different scenarios, for example, in the context of multimode fields \cite{walborn2003multimode}, with $^4{\rm He}$ atoms \cite{lopes2015atomic}, in the plasmonic regime \cite{di2014observation}, in the magnonic regime \cite{kostylev2023magnonic}, with microwave fields \cite{lang2013correlations}, in atomic ensembles \cite{li2016hong}, in four-wave mixing \cite{chen2015measuring}, with different color photons \cite{kobayashi2016frequency}, and on photonic chips with topologically protected states \cite{tambasco2018quantum}.

\begin{figure}
\centering
    \includegraphics[width=2.8in, height=2.75in]{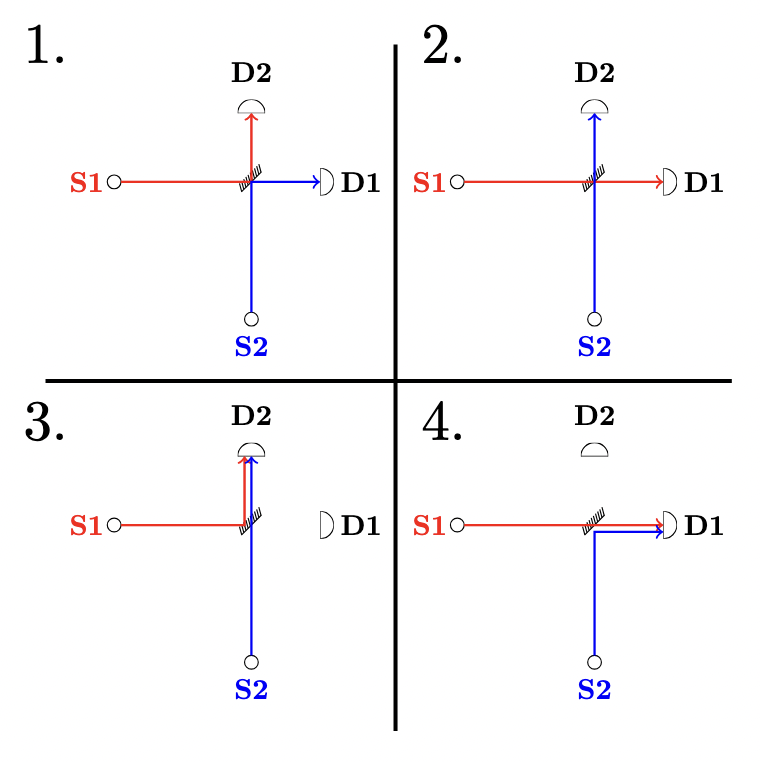}
\captionsetup{
  format=plain,
  margin=1em,
 justification=raggedright,
 singlelinecheck=false
}
\caption{(Color online) The four possible outcomes in the standard HOME setup. }\label{Fig1}
\end{figure}

The Hong-Ou-Mandel effect in its standard setting can be introduced in the following fashion \cite{branczyk2017hong, bouchard2020two}. Consider two-photon sources (S1, S2) arranged against a perfect $50/50$ beam splitter, as shown in Fig.~\ref{Fig1}. We place two ideal photodetectors, D1 and D2, at the two output ports. Then, the four classical outcomes (each representing a different combination of counts at the detectors) can be drawn, as shown in Fig.~\ref{Fig1}.  
What Hong, Ou, and Mandel were able to demonstrate was the fact that the two indistinguishable photons arriving at the same time at the input ports of such a beam splitter will always emerge together at the two output detectors (i.e., only the two lowest outcomes shown in Fig.~\ref{Fig1} will ever be observed \cite{hong1987measurement}). Furthermore, they found that photons incident on the beam splitter within a very small time interval have a statistically significant likelihood of being detected at the same detector, with the likelihood of this correlation's occurrence decreasing as the interval of incidence increased. This gives rise to the well-known HOME dip \cite{gerrits2015spectral}. The final state at the two detectors $\frac{1}{\sqrt{2}}(\ket{2,0}+\ket{0,2})$ thus takes the simplest possible form in the much larger class of quantum states of light called the ${\rm N00N}$ states \cite{kok2002creation} which have their applications in different areas of quantum information science (for applications in quantum metrology, see Ref.~\cite{dowling2008quantum}).

Most standard mathematical treatments of HOME only specify the presence of single-photon sources without detailing the actual physical mechanism responsible for generating single photons. Since the reliable generation of indistinguishable single photons is a key element of the HOME, in this work, we address this problem by considering two special cases: (1) when the single photons are generated by two initially excited two-level atoms and (2) two atom-cavity systems which under the standard Jaynes-Cummings like interaction \cite{larson2021jaynes} are known to be capable of generating single photons \cite{dayan2008photon}. To understand the entire process of photodetection, we employ the quantum jump/trajectory approach in the same spirit as introduced by Carmichael in his classic references \cite{carmichael2009open, carmichael1993quantum}. To this end, we use both analytical and numerical methods. Our analytical approach relies on the time evolution of the system under a non-Hermitian Hamiltonian, stochastically interrupted by quantum jumps, which account for the detection of single photons by the output detectors. Our numerical treatment relies on mimicking actual photodetection events using a Monte Carlo simulation based on the quantum jump/trajectory approach.

As some of the main results of this work, we find that in the excited atom source case, the detection of the first photon maps the initial state of the system to an entangled Bell-like state. For identical atoms, the detection of the second photon never occurs at a different detector, confirming perfect homodyne detection (HOME). The same detector probability is controlled by the atomic decay rate parameter, indicating the impact of the single-photon source utilized. We find that these analytic results extend to atom-cavity single-photon sources as well, where we observe the presence of a perfect HOME effect with post-jump state dependent on the atom-cavity coupling strength. Next, in our Monte Carlo simulation results, we observe that the excitation probabilities in both atom-cavity systems exhibit Rabi oscillations in the strong-coupling regime of the cavity QED. Furthermore, we computed coincidence percentages and found the departure from the perfect HOME, as atom-cavity systems tend to emit photons with different cavity leakage rates.

The remainder of the manuscript is structured as follows. In Sec.~\ref{sec:II}, we present our analytic results for the excited atom sources. Next, we devote the first part of Sec.~\ref{sec:III} to the analytic results for atom-cavity sources. In the second part of this section, we discuss our numerical results for different parameter regimes. In the last part of Sec.~\ref{sec:III}, we study the time delay between the counts for different parameter regimes and compare our results of coincidence and anticoincidence for independent atom-cavity sources. Finally, we conclude with Sec.~\ref{sec:IV} by summarizing our main results. 

%%===================================================%%
%%          Sec.II: HOMI with Excited Atoms          %%
%%===================================================%%
\section{\label{sec:II} HOME with Excited Atom Sources}
\begin{figure}
\centering
    \includegraphics[width=2.7in, height=2in]{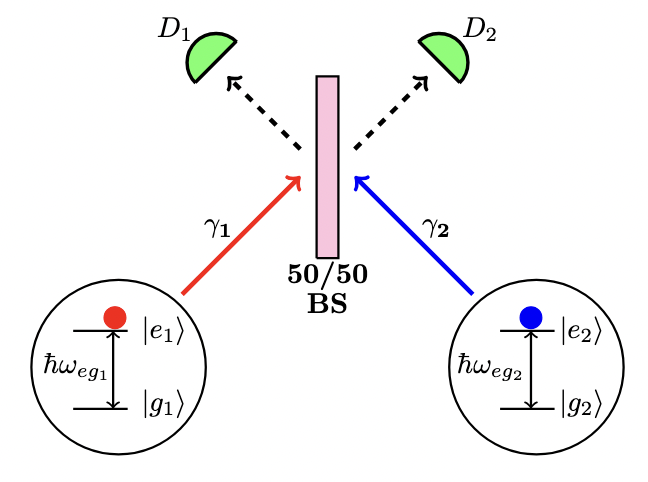}
\captionsetup{
  format=plain,
  margin=1em,
 justification=raggedright,
 singlelinecheck=false
}
\caption{(Color online) The diagram of the two-photon Hong-Ou-Mandel effect with two excited atoms serving as single photon sources. $\gamma_1$ and $\gamma_2$ are the decay rates with which the photons are randomly emitted and arrive at the input ports of a $50/50$ beam splitter. $D_1$ and $D_2$ are perfect detectors placed at the two output ports.}\label{Fig2}
\end{figure}

\begin{figure*}
\centering
    \includegraphics[width=4.75in, height=0.7in]{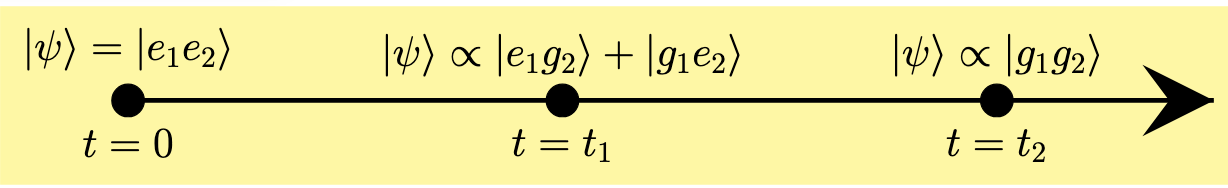}
\captionsetup{
  format=plain,
  margin=1em,
 justification=raggedright,
 singlelinecheck=false
}
\caption{(Color online) A timeline summarizing the main steps of the quantum jump protocol. The system starts with both atoms being excited. At a later random time $ t_1$, one of the two detectors clicks, triggering the collapse of our system and the formation of a Bell state between the two atoms. The process concludes with the other clicks occurring randomly at time $t_2>t_1$, leading to both jumps being recorded at that time and the system relaxing to the ground state.}\label{Fig3}
\end{figure*}

%\subsection{\label{sec:IIA} Model Hamiltonian}
We begin the main body of our paper by discussing HOME, utilizing excited atoms as single-photon sources. As shown in Fig.~\ref{Fig2}, we express the Hamiltonian of the system under consideration, which consists of two independent two-level atoms in an excited state:
\begin{align}\label{sysHam}
    \hat{\mathcal{H}}_{sys} = \hbar\omega_{eg_{1}}\hat{\sigma}^\dagger_1\hat{\sigma}_1 + \hbar\omega_{eg_{2}}\hat{\sigma}^\dagger_2\hat{\sigma}_2,
\end{align}
where $\hbar\omega_{eg_j}$ is the atomic transition frequency and $\hat{\sigma}_j:=\ket{g_j}\bra{e_j}$ the atomic lowering operator for the $j$th atom. Here, for simplicity, we set the ground-state energy of the atoms to zero. The nonvanishing anticommutation relation between the atomic operators is given by: $\lbrace \hat{\sigma}_j, \hat{\sigma}^\dagger_k \rbrace = \delta_{jk}$, $\forall$, $j=1,2$ and $k=1,2$.
As shown in Fig.~\ref{Fig3}, the system starts off with an initial state 
\begin{align}
\ket{\psi(t=0)} = \ket{e_1}\otimes\ket{e_2}\equiv \ket{e_1e_2}. 
\end{align}
Using the Collett and Gardiner input-output theory \cite{gardiner1985input}, the quantum jump operator for the first atom $\hat{J}_1$ can be constructed to be 
\begin{align}
\hat{J}_1 = \hat{\sigma}_{in} + \sqrt{\gamma_1} \hat{\sigma}_1,
\end{align}
where $\hat{\sigma}_{in}$ represents the input operator for the first atom. Since no such input is present in our model, we disregard these operator terms in the analysis that follows. Similarly, for the second atom, we write the jump operator as: $\hat{J}_2 = \sqrt{\gamma_2}\hat{\sigma}_2$. The $50/50$ beam splitter mixes the two operators, producing two newly transformed operators $\hat{J}_{\pm}$, which are given by
\begin{align}\label{jumpop1}
    \hat{J}_{\pm} = \frac{\hat{J}_1 \pm \hat{J}_2}{\sqrt{2}} = \sqrt{\frac{\gamma}{2}}\left( \hat{\sigma}_1\otimes\hat{1}_2 \pm \hat{1}_1\otimes \hat{\sigma}_2\right).
\end{align}
Here, for simplicity, we have assumed that spontaneous emission rates for both atoms are the same and $\hat{1}_j$ represents the identity operator that indicates the completeness relation belonging to the Hilbert space for the $j$th atom. According to the quantum jump approach \cite{carmichael2009open, dalibard1992wave, daley2014quantum}, before the detection of any photon and between any two consecutive jumps, a quantum system evolves according to a non-Hermitian Hamiltonian operator. In the present case, this operator is constructed from the Hermitian system Hamiltonian (as given by Eq.~\eqref{sysHam}) and beam splitter quantum jump operators defined in Eq.~\eqref{jumpop1} as
\begin{align}
    \hat{\mathcal{H}}_{NH} = \hat{\mathcal{H}}_{sys} - \frac{i\hbar}{2}\sum_{j=\pm}\hat{J}^\dagger_{j}\hat{J}_{j} = \sum^2_{i=1}\left(\hbar\omega_{eg_{i}} - \frac{i\hbar}{2}\gamma_i\right)\hat{\sigma}^\dagger_i\hat{\sigma}_i.
\end{align}
Next, we define time $t_1$ as the time when, stochastically, the first jump occurs (as indicated by clicking on one of the two detectors). Just before $t=t_1$, our initial state would evolve to take the following form:
\begin{align}
  \ket{\psi(t<t_1)} = e^{-i\hat{\mathcal{H}}_{NH}t/\hbar}  \ket{\psi(0)} = e^{-2i\omega_{eg}t}e^{-\gamma t}\ket{e_1e_2}.
\end{align}
In the last equation, for the sake of simplicity, we have assumed that both atoms are identical, i.e. $\omega_{eg_{i}} = \omega_{eg}$ and $\gamma_i = \gamma$, $\forall i=1,2$. Now, suppose that at time $t=t_1$, one of the two photodetectors clicks. If this click is caused by the jump operator $\hat{J}_{+}$, then according to quantum jump theory, the renormalized state of our quantum system at that time will be found to be
\begin{align}
    \ket{\psi(t_1)} \longrightarrow &\frac{{\hat{J}_{+}\ket{\psi(t)}}}{\sqrt{\bra{\psi(t_1)}\hat{J}^\dagger_{+}\hat{J}_{+}\ket{\psi(t_1)}}} \nonumber\\
    &= \frac{e^{-2i\omega_{eg}t_1} }{\sqrt{2}}\left(\ket{g_1e_2}+\ket{e_1g_2} \right).
\end{align}
The denominator in the above equation represents the normalization factor. In Fig.~\ref{Fig3}, we describe the evolution of our system as it undergoes different stages of the quantum jump protocol. It is remarkable to note that just after the recording of the first jump, the two atoms form a maximally entangled Bell state \cite{brunner2014bell, yang2018entanglement} with the atomic transition frequency (a parameter characterizing our single-photon source) appearing in the global phase factor. 

Afterward, the system again evolves according to the non-Hermitian Hamiltonian until the next jump is recorded. At $ t=t_2 $ is the time that the second (and last) jump is recorded. Thus, for the time interval $t_1<t<t_2$, the state of the system takes the form
\begin{align}
    \ket{\psi(t<t_2)} = e^{-i\omega_{eg}(t+t_1)}e^{-\gamma (t-t_1)/2}\frac{\left(\ket{g_1e_2}+\ket{e_1g_2} \right)}{\sqrt{2}}.
\end{align}
We note that this state is still a Bell-like state. However, in addition to a global phase factor that depends on the atomic transition frequency $\omega_{eg}$, there is now another exponential decay factor that depends on the atomic decay rate $\gamma$. Finally, at some later time $t_2$, one of the two detectors clicks. As a result, both atoms were found to be in their ground state, thereby concluding the single realization of our jump protocol. If the first jump was caused by (say) the operator $\hat{J}_{+}$, there are two possibilities in the recording of the second jump:
\begin{itemize}
    \item In the first scenario, our second jump is caused by the same jump operator $\hat{J}_{+}$, in which case the state of our quantum system will be found to be
    \begin{align}
    \ket{\psi(t_2)} \longrightarrow & \frac{{\hat{J}_{+}\ket{\psi(t)}}}{\sqrt{\bra{\psi(t_2)}\hat{J}^\dagger_{+}\hat{J}_{+}\ket{\psi(t_2)}}} \nonumber\\
    &= \sqrt{\gamma}e^{-i\omega_{eg}(t_1+t_2)}e^{-\gamma(t_2-t_1)/2}\ket{g_1g_2},
    \end{align}
    with the corresponding probability (let's call it $P_{+}$) in some tiny time interval $dt\ll\gamma^{-1}$ given by
    \begin{align}
        P_{+}(t_2)dt = \bra{\psi(t_2)}\hat{J}^\dagger_+\hat{J}_+ \ket{\psi(t_2)}dt = \gamma e^{-\gamma(t_2-t_1)}dt.
    \end{align}
    \item In the second scenario, let us consider that the second jump is caused by the operator $\hat{J}_{-}$, which results in the probability (let us call it $P_{-}$) in some tiny time interval $dt$ given by
    \begin{align}
        P_{-}dt = \bra{\psi(t_2)}\hat{J}^\dagger_-\hat{J}_- \ket{\psi(t_2)}dt = 0.
    \end{align}
\end{itemize}
Following the standard literature on HOME \cite{branczyk2017hong}, both of these results can be combined into a single quantity known as the visibility parameter $\mathcal{V}$ \cite{close2012overcoming, nazir2009overcoming}. For the present quantum jump/trajectory protocol, visibility takes the following form
\begin{align}
    \mathcal{V} = \left| \frac{P_{\text{same}} - P_{\text{diff}}}{P_{\text{same}} + P_{\text{diff}}}\right|,
\end{align}
where $P_{\text{same}} = P(J_+|J_+) + P(J_-|J_-)$ and $P_{\text{diff}} = P(J_+|J_-) + P(J_-|J_+)$ with $P(J_x|J_y)$ being the probability of detecting one jump with the $\hat{J}_x$ operator given that the first jump has occurred due to the application of the $\hat{J}_y$ operator. In the above analysis, we first considered the operator $\hat{J}_+$ and then $\hat{J}_+$ or $\hat{J}_-$ was applied. We found that either this order of operators is followed or the other case in which $\hat{J}_-$ acts first and then either $\hat{J}_+$ or $\hat{J}_-$ is applied, the visibility parameter $\mathcal{V}$ turns out to take a unit value that shows the occurrence of a perfect HOME for the two-excited atom single-photon source case.  

%%===================================================%%
%%  Sec.III Single emitter-cavity problem            %%
%%===================================================%%
\section{\label{sec:III} HOME With Cavity QED Sources}
\subsection{Analytic results}
\begin{figure}
\centering
    \includegraphics[width=3.5in, height=2.3in]{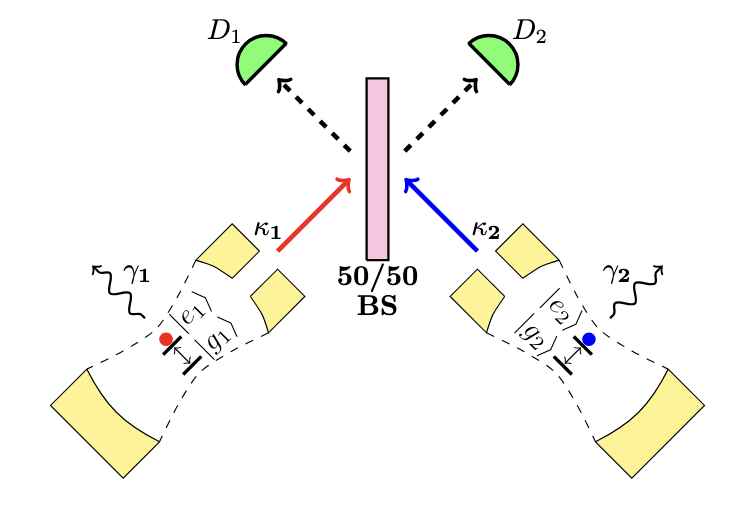}
\captionsetup{
  format=plain,
  margin=1em,
 justification=raggedright,
 singlelinecheck=false
}
\caption{(Color online) The experimental setup, as depicted in Fig.~\ref{Fig1}, is the same as in our model here, with a key difference: now, single photons are generated by cavity QED interactions. The parameter $\kappa_j$ indicates the photon leakage rate from the $j$th cavity, and $\gamma_j$ represents the spontaneous emission rate for the $j$th atom. In our model, we suppose that the cavity leakage rate is the process responsible for providing input photons to the 50/50 beam splitter. This is in stark contrast to spontaneous emission processes, which represent another channel that causes the pure loss of energy from the system. }\label{Fig4}
\end{figure}

Now, we turn our attention to the problem of HOME with single photons that are generated by cavity quantum electrodynamics (or cavity QED) sources. In particular, we ask this question: {\it Do the single-photons generated by cavity QED sources in the strong or weak coupling regime leave any imprints on the two-photon interference?} To this end, as shown in Fig.~\ref{Fig4}, we assume that the photon sources are Fabry-Pérot cavities, each containing a two-level atom described as being in either the excited state $\ket{e_j}$ or the ground state $\ket{g_j}$ (with $j=1,2$), coupled to the single mode of an electromagnetic field which has been truncated to one excitation at maximum. One possible way of realizing such a situation would be to weakly excite a pair of identical Fabry-Pérot cavities with a coherent drive (i.e., a low-intensity laser beam). Previous work \cite{liang2019antibunching} has indicated that such a setup may utilize photon blockade for the emission of single photons.

The initial state of the system is the tensor product of each atom's ground state and the associated field's Fock state, which initially contains a single photon, as given by the following equation
\begin{align}
    \ket{{\psi}(t=0)} = \ket{g_1} \otimes \ket{g_2} \otimes \ket{1} \otimes \ket{1}\equiv \ket{g_1g_2;11}.
\end{align}
Again, we employ the quantum jump approach \cite{carmichael2009open} as our primary theoretical tool to investigate the stochastic time evolution of our initial state under the given model. To this end, we introduce two jump operators, $\hat{J}_+$ and $\hat{J}_-$, to represent the detection of a photon at Detector 1 and Detector 2, respectively. Each jump operator is constructed from the cavity field operators under the beam-splitter transformation as
\begin{align}
    \hat{J}_{\pm} = \sqrt{\kappa}\left(\frac{\hat{a}_{1} \pm \hat{a}_{2}}{\sqrt{2}}\right),
\end{align}
where $\kappa$ is the photon leakage rate, which, for simplicity, we have assumed to be the same for both cavities, and $\hat{a}_j$ is the photon annihilation operator for the $j$th cavity. To account for the time evolution of the state in between successive quantum jumps, we define the non-Hermitian Hamiltonian for the present setup under the interaction picture as follows:
\begin{figure*}
\centering
    \includegraphics[width=5.25in, height=0.85in]{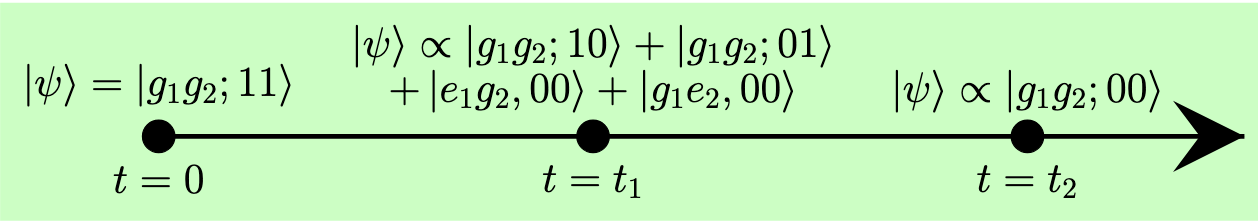}
\captionsetup{
  format=plain,
  margin=1em,
 justification=raggedright,
 singlelinecheck=false
}
\caption{(Color online) Quantum jump-based evolution timeline outlining the different steps involved in the protocol for cavity QED-based single photon sources. Note that, except for the initial state, all intermediate quantum states appear with time-dependent functions multiplied by different terms in the states. We have only indicated the state parts above for simplicity.}\label{Fig5}
\end{figure*}
\begin{align*}
&\hat{\mathcal{H}}_{NH} = \sum^2_{j=1}\left[\hbar G_j(\hat{a}_j^\dagger \hat{\sigma}_j + \hat{\sigma}_j^\dagger \hat{a}_j)\right] \nonumber\\
&- \frac{i\hbar}{2}\hat{J}^\dagger_+\hat{J}_+ - \frac{i\hbar}{2}\hat{J}_-^\dagger\hat{J}_- -\frac{i\hbar\gamma}{2}\hat{\sigma}_1^\dagger\hat{\sigma}_1 - \frac{i\hbar\gamma}{2} \hat{\sigma}_2^\dagger \hat{\sigma}_2.
\end{align*}
Here, $G_j$ is the atom-field interaction parameter for the $j^{th}$ atom-cavity system. $\hat{\sigma}_j := \ket{g_j}\bra{e_j}$ is the $j^{th}$ atom lowering operator. The ladder operators for the photons in the optical cavity follow the Bosonic commutation relation $[\hat{a}_i, \hat{a}^\dagger_j] = \delta_{ij}$. For the Hermitian part of the above Hamiltonian, one can find the time-evolution operator between some initial time $t_i$, and some final time $t_f$ becomes \cite{scully1997quantum}:
\begin{widetext}
\begin{align}
  \hat{U}(t)= e^{\frac{i\hat{\mathcal{H}}(t_f-t_i)}{\hbar}} &=\prod\limits^2_{j=1}\Bigg[\cos{\left(G_j t\sqrt{\hat{N}_j+1}\right)}|e_j\rangle\langle e_j| + \cos{\left(G_j t\sqrt{\hat{N}_j}\right)}|g_j\rangle\langle g_j|\nonumber\\
  & -\frac{i}{\sqrt{\hat{N}_j+1}}\left\lbrace\sin{\left(G_j t\sqrt{\hat{N}_j+1}\right)}\hat{a}_j|e_j\rangle\langle g_j| +\hat{a}_j^{\dagger}\sin{\left(G_j t\sqrt{\hat{N}_j+1}\right)}|g_j\rangle\langle 
e_j|\right\rbrace\Bigg],
\end{align}
\end{widetext}
with the definition introduced $\hat{N}_j:=\hat{a}_j^\dagger\hat{a}_j$. As indicated in the timeline shown below (see Fig.~\ref{Fig5}), the quantum jump protocol obeys the following steps:
\begin{itemize}
    \item The protocol begins with an initial state $\ket{\psi(t=0)} = |g_1 g_2; 11\rangle$. Once the evolution starts, then these single photons can either be absorbed by the atoms through cavity QED interactions (where later, at some random time, these atoms can emit the photons back into their respective cavities or lose the photons through spontaneous emission channel) or photons can be leaked out from the cavity and arrive at the input ports of the beam splitter. 
    \item This initial state evolve from $t=0$ to $t_1$ under no-photon detection constraint by applying the time-evolution operator: $|\psi(t_1)\rangle = \hat{U}(t)|\psi(t=0)\rangle$. After lengthy but straightforward calculations, we find the state just before the time $t_1$ takes the form
    \begin{align}
        & \ket{\psi(t<t_1)} = ~ e^{-\kappa t_1}\cos^2(Gt)\ket{g_1 g_2; 1 1} \nonumber\\
        &- i e^{-(\frac{\kappa}{2} + \frac{\gamma}{4})t_1} \cos(Gt)\sin(Gt)\Big(\ket{e_1 g_2; 0 1}+\ket{g_1 e_2; 1 0}\Big) \nonumber\\
        & - e^{-\frac{\gamma}{2}t_1}\sin^2(Gt)\ket{e_1 e_2; 0 0},
    \end{align}
    where we have assumed both optical cavities and atoms are identical. We observe that the inclusion of the anti-Hermitian terms manifests as multiplicative decaying terms, dependent on the loss parameters $\kappa$ and $\gamma$. Furthermore, we point out that for the cavity QED single photon sources, the atom-cavity coupling rate $G$ (needed to distinguish between strong and weak coupling regimes of cavity QED in terms of the cooperativity parameter $\mathcal{C}=2G^2/(\kappa\gamma)$ \cite{larson2021jaynes,zhou2023coupling}) begins to enter in our results.
    \item At some arbitrary time, $t_1$, a jump randomly occurs due to the leakage of a photon from one of the cavities. As we lack information about \textit{which} cavity emitted the photon, we express the resultant state as a superposition of four possibilities: $\ket{\psi(t_1)} \propto |g_1g_2; 10\rangle+|g_1g_2; 01\rangle + \ket{g_1 e_2; 0 0} + \ket{e_1 g_2; 0 0}$. More specifically, we find that for the case of applying $\hat{J}_+$, the resultant state post the detection of this first quantum jump can be expressed as
    \begin{align}
        & \ket{\psi(t_1)} =  e^{-\kappa t_1}\cos^2(Gt_1) \frac{\ket{g_1 g_2; 1 0} + \ket{g_1 g_2; 0 1}}{\sqrt{2}}\nonumber\\
        & - ie^{-(\frac{\kappa}{2} + \frac{\gamma}{4})t_1}\cos(Gt_1)\sin (Gt_1)\frac{\ket{e_1g_2; 00} + \ket{g_1e_2; 00}}{\sqrt{2}}.
    \end{align}
    \begin{figure*}
\centering
    \includegraphics[width=2.61in, height=1.85in]{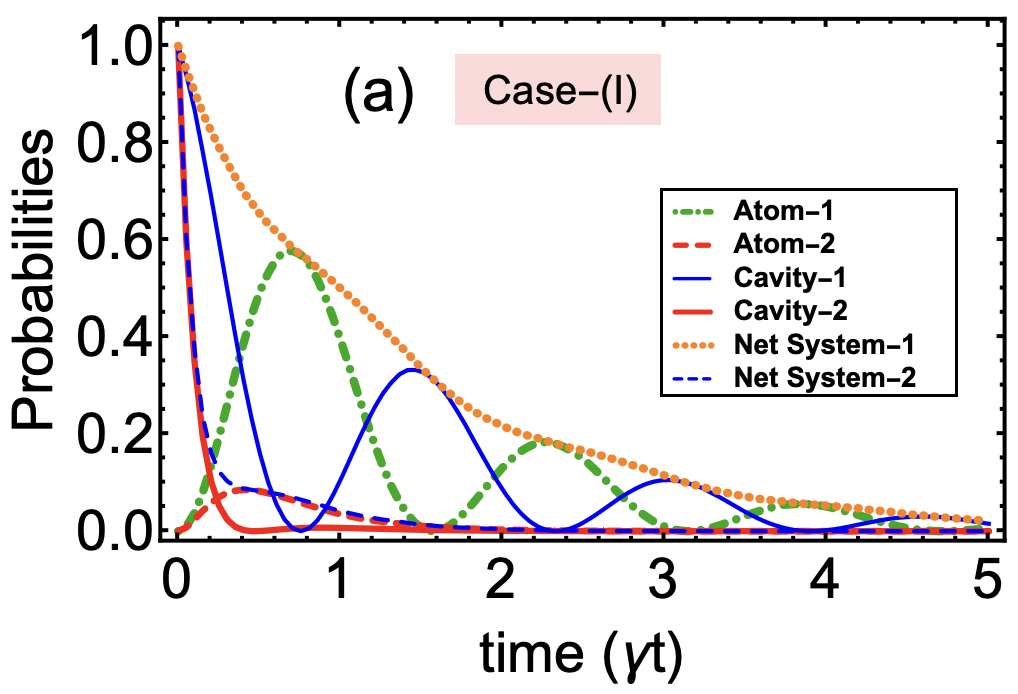}
    \includegraphics[width=2.61in, height=1.85in]{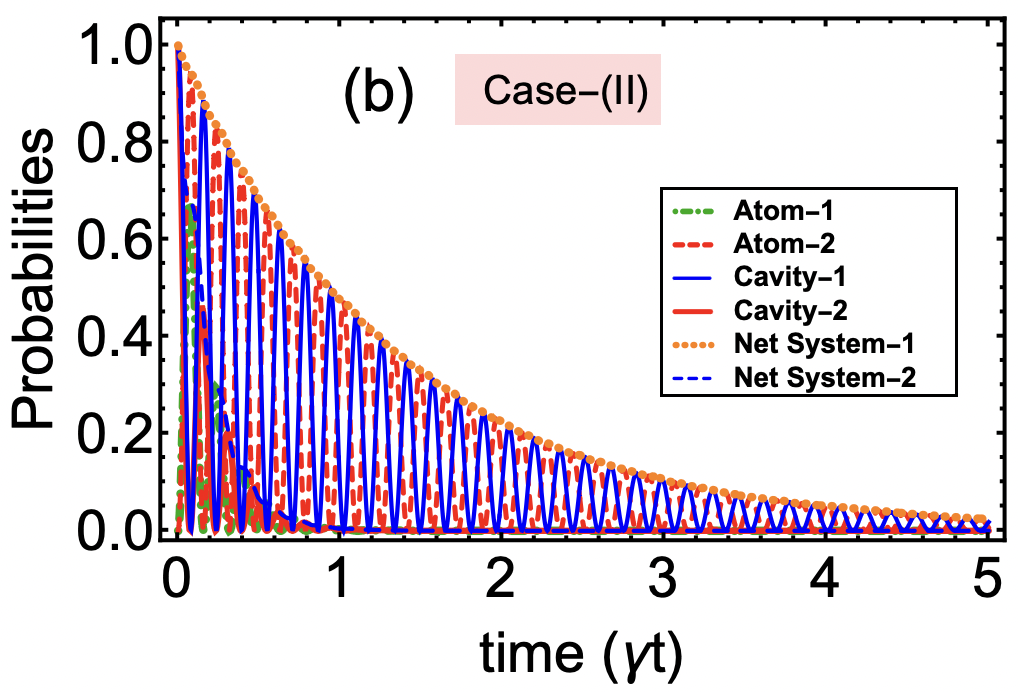}
\captionsetup{
  format=plain,
  margin=1em,
 justification=raggedright,
 singlelinecheck=false
}
\caption{(Color online) Expectation values of the number operator and excitation operators as a function of time (in the units of $\gamma$) with $\gamma_1=\gamma_2=\kappa_1=\gamma$ and $\kappa_2=10\gamma$ while in (a) Case-(I): $g_1=g_2=2\gamma$ and in (b) Case-(II): $g_1=g_2=20\gamma$. In both cases, the averages have been calculated over $10,000$ trajectories with errors too small to be displayed on the curves.}\label{Fig6}
\end{figure*}
    After that, we normalize the above state, which, as shown in Fig.~\ref{Fig5}, yields
    \begin{align}
    \ket{\psi(t_1)} = & \frac{1}{\sqrt{2}}\Big[\cos (Gt_1) \Big(\ket{g_1 g_2; 1 0} + \ket{g_1 g_2; 0 1}\Big) \nonumber\\
    & - i\sin (Gt_1) \Big(\ket{g_1 e_2; 0 0} + \ket{e_1 g_2; 0 0}\Big)\Big].
    \end{align}
    \item Next, a quantum jump occurs arbitrarily at time $t=t_2$. Up to time $t_2$, we evolve the state once again, using the evolution equation $|\psi(t<t_2)\rangle = \hat{U}(t-t_1)|\psi(t_1)\rangle$. As a result, we find
    \begin{align}
        &\ket{\psi(t<t_2)} = \frac{1}{\sqrt{2}}\Bigg[e^{-\frac{\kappa \Delta t}{2}}\Big(\cos (Gt)\cos (G\Delta t)\nonumber\\
        &-\sin(Gt)\sin(G\Delta t)\Big)\Big(\ket{g_1 g_2; 1 0} + \ket{g_1 g_2; 0 1}\Big) \nonumber\\& - ie^{-\frac{\gamma \Delta t}{4}}\Big(\sin (Gt)\cos(G \Delta t) + \cos (Gt)\sin(G\Delta t)\Big)\times\nonumber\\
        &\Big(\ket{e_1 g_2; 0 0} + \ket{g_1 e_2; 0 0}\Big)\Bigg],
    \end{align}
   where we have adopted a short notation in which $\Delta t \equiv t - t_1$.
   \item Finally, at time $t_2$, the occurrence of a random jump results in a second click at one of the two detectors. To represent this situation, we apply our second jump operator (again choosing between either $\hat{J}_+$ or $\hat{J}_-$). For example, in the case of applying $\hat{J}_-$ operator, we find 
   \begin{align}
     &\hat{J}_-\ket{\psi(t_2)} = \sqrt{\frac{\kappa}{2}}\Big(e^{-\frac{\kappa \delta t}{2}}(\cos (Gt_1)\cos (G\delta t)\nonumber\\
     &-\sin(Gt_1)\sin(G\delta t)\Big)\Bigg(\frac{\ket{g_1 g_2 0 0}}{\sqrt{2}} + 0 -0 - \frac{\ket{g_1g_200}}{\sqrt{2}}\Bigg)\nonumber\\
     & - ie^{-\frac{\gamma \delta t}{4}}\Big(\sin (Gt_1)\cos(G \delta t) + \cos (Gt_1)\sin(G\delta t)\Big)(0)),
   \end{align}
   where $\delta t=t_2 - t_1$. On the other hand, if we apply the $\hat{J}_+$ the second time as well, we find the final state takes the form 
   \begin{align}
       &\hat{J}_+\ket{\psi(t_2)} = \sqrt{\kappa}e^{-\frac{\kappa \delta t}{2}}\Big(\cos(Gt_1)\cos(G\delta t)\nonumber\\
       &-\sin(Gt_1)\sin(G\delta t)\Big)\ket{g_1 g_2 0 0}.
   \end{align}
\end{itemize}
We conclude that the second jump with $\hat{J}_-$ applied conditioned on the fact that the first jump was caused by $\hat{J}_+$ yields a zero probability. On the other hand, if both jumps are caused by the same operator (either by the jump operator $\hat{J}_+$ or by the operator $\hat{J}_-$), we are left with the state $\ket{\psi(t_2)} \propto |g_1g_2,00\rangle$. This shows the presence of a perfect HOME for QED-based single-photon cavity sources with the final state (for the nonzero outcome case) dependent on the single-photon source, namely, the cavity leakage rate $\kappa$ and the atom-cavity coupling strength parameter $g$. In the following subsection, we move on to discuss our numerical results for the cavity QED photon sources.

\subsection{\label{sec:IIIA} Numerical results: Monte Carlo simulations}
To perform a numerical study of our single-photon atom-cavity sources, we apply the Monte Carlo simulation based on the quantum jump/trajectory approach \cite{lambert2024qutip}. Overall, we find excellent agreement between our numerical and analytic results, with numerical results providing additional insights into our main conclusions. To begin with, we computed the time-dependent expectation values of the number operator $\hat{a}^\dagger_n\hat{a}_n$, the atomic excitation operator $\hat{\sigma}^\dagger_n \hat{\sigma}_n$ and the total excitation operator $\sum_n(\hat{a}^\dagger_n\hat{a}_n + \hat{\sigma}^\dagger_n\hat{\sigma}_n)$ for the $n^{th}$ atom-cavity source (here $n=1,2$). 

\begin{figure}
\centering
    \includegraphics[width=2.5in, height=1.7in]{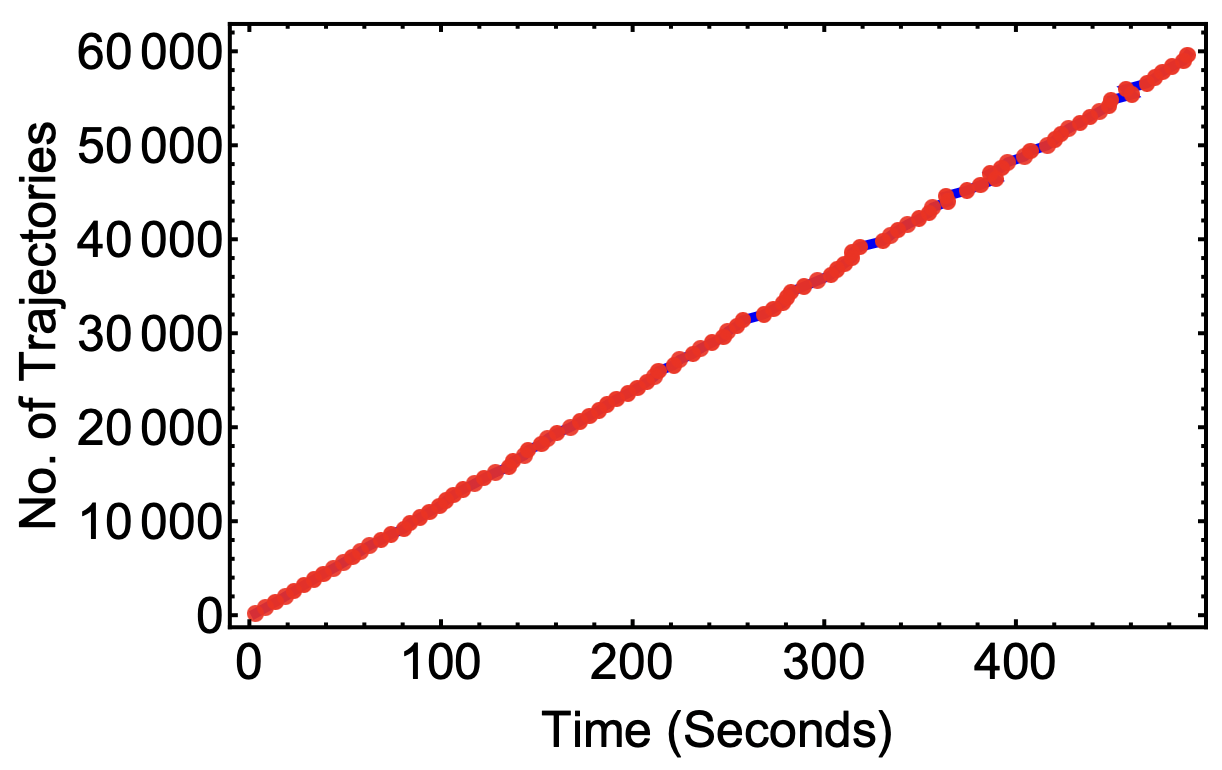}
\captionsetup{
  format=plain,
  margin=1em,
 justification=raggedright,
 singlelinecheck=false
}
\caption{(Color online) A plot showing the dependence of the computing time on the number of trajectories for our Monte-Carlo simulation to generate numerical results that are reported here.}\label{Fig7}
\end{figure}

Fig.~\ref{Fig6} shows the results of our simulations for two cases. In the first case (Fig.~\ref{Fig6}(a)), we consider the cooperativity for the first and second atom-cavity systems to be $\mathcal{C}_1:= 2g^{2}_1/(\kappa_1\gamma_1)=8$ and $\mathcal{C}_2=0.8$, respectively. This corresponds to a strong (weak) coupling regime for the first (second) atom-cavity system. In this case, we observe the presence of damped Rabi oscillations for the first atom-cavity system, indicating the back-and-forth absorption and emission of photons between the atom and the cavity. For the second atom-cavity system, due to the weak coupling regime, the Rabi oscillations are washed out, and excitation probabilities exhibit purely decaying behavior. For both first- and second-atom-cavity systems, we find that the net excitation probability (see orange dotted and blue dashed curves) bounds the sum of the individual atomic and cavity excitation probability as expected. Finally, in Fig.~\ref{Fig6}(b) we considered strong coupling regimes for both atom-cavity systems with $\mathcal{C}_1=800$ and $\mathcal{C}_2=80$. As expected, with such high cooperativity values (with $\mathcal{C}_2$ experimentally feasible \cite{zhou2023coupling} but $\mathcal{C}_1$ being a much futuristic possibility), we obtain extremely rapid Rabi oscillations, which still are damped, and the net excitation probability posing an upper bound on individual probability for both atom-cavity subsystems. 

\begin{figure*}
\centering
    \includegraphics[width=2.5in, height=1.7in]{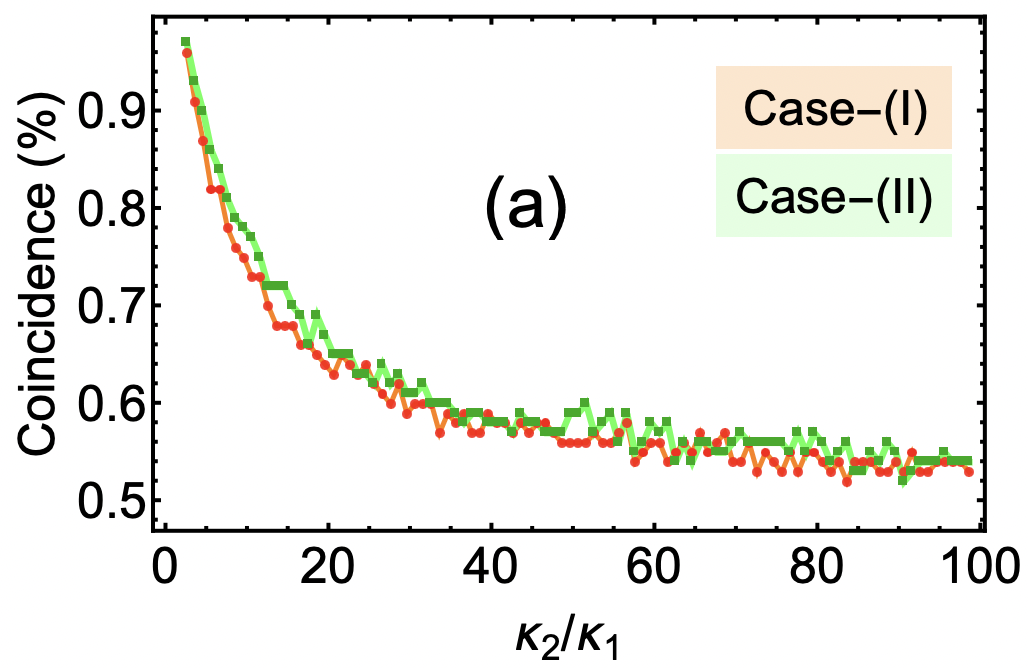}
    \includegraphics[width=2.5in, height=1.7in]{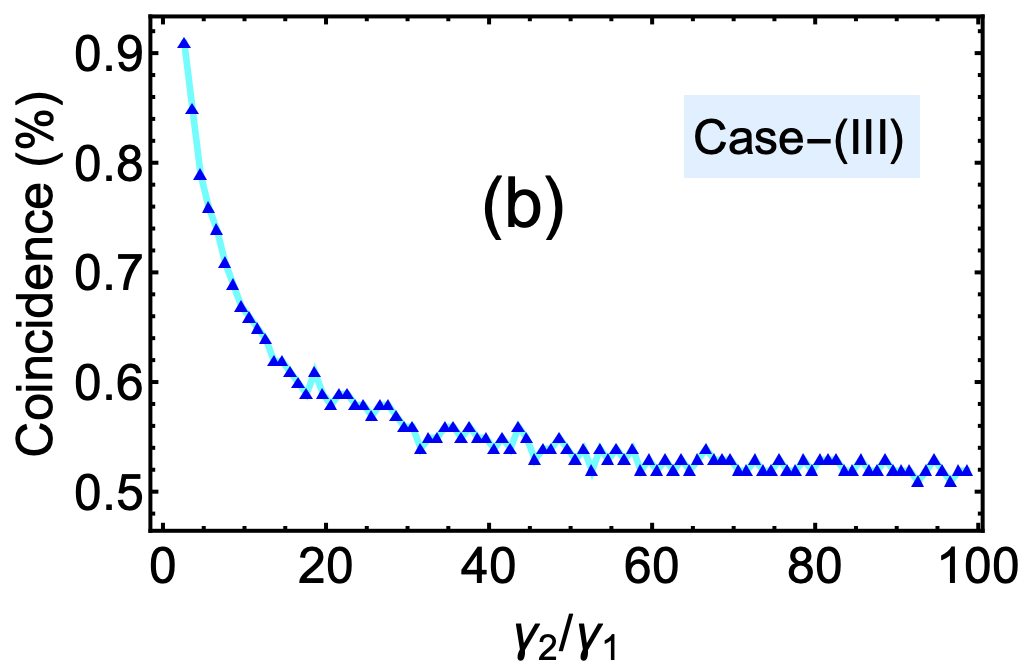}
\captionsetup{
  format=plain,
  margin=1em,
 justification=raggedright,
 singlelinecheck=false
}
\caption{(Color online) Coincidence percentage for (a) atom-cavity- and (b) case-III atom-based single photon sources. In plot (a), for both cases, we have set $\kappa_1=\gamma_1=\gamma_2=\gamma$ and varied $\kappa_2$ in terms of $\kappa_1$ up to $100\kappa_1$, while for case (I) $g_1=g_2=\gamma$ and for case (II) $g_1=g_2=10\gamma$. For plot (b), on a similar note, we have fixed $\gamma_1=\gamma$ and varied $\gamma_2$ in terms of $\gamma_1$ up to $100\gamma_1$.}\label{Fig8}
\end{figure*}

Before proceeding to our coincidence count results, we report an interesting numerical detail: the dependence of the number of trajectories on the time required for our Monte Carlo simulation to complete. The resultant data is presented in Fig.~\ref{Fig7}, where we find a positive linear relation between the two variables. Given that the marginal cost is constant, we based our decision on the desired error. We found that using $10\cdot 10^4$ trajectories allowed us to minimize our error to $0.316\%$, which is too tiny an error to be displayed on the numerical curves presented in this work.

We now report our results on the impact of different cavity QED-based single-photon sources on two-photon interference. To this end, we assume that the cavity leakage rates of the two atom-cavity systems differ, and we use the coincidence count percentage to quantify the deviation from a perfect HOME. Therein, a coincidence of 1 means a 100\% probability of detecting both photons at the same detector, and a coincidence of 0.5 indicates a complete loss of the interference phenomenon. We report these results in Fig.~\ref{Fig8}. In Fig.~\ref{Fig8}(a), we plot two cases of parameters on the same scale. For case (I), we have set $g_1=g_2=\gamma_1=\gamma_2=\gamma$ and $\kappa_1=\gamma$ while varying $\kappa_2$, while for case (II) all parameters are the same except for $g_1=g_2=10\gamma$. For comparison with the excited-atom problem of Sec.~\ref{sec:II}, we have also included Fig.~\ref{Fig8}(b), where $\gamma_2/\gamma_1$ is used as a ratio to quantify the difference between two excited-atom-based photon sources. 

In Fig.~\ref{Fig8}(a), we note that both cases start from a perfect coincidence when $\kappa_2=\kappa_1$. After that, as $\kappa_1$ and $\kappa_2$ begin to become different, the coincidence percentage decreases with case (II) (stronger coupling regime compared to case (I)), remaining slightly higher than the coincidence curve in case (I). We also notice a somewhat oscillatory behavior, more prominent for case (II), later in the graph (see, for instance, around $\kappa_2=50\kappa_1$). For the two-atom sources of Fig.~\ref{Fig8}(b), we notice the same overall trend, with the difference that the decay in the coincidence is faster due to the lack of atom-cavity interactions. Finally, we observe that in both plots, the coincidence percentage asymptotically approaches fifty percent as $\kappa_2/\kappa_1\gg1$ or $\gamma_2/\gamma_1\gg1$. We explain this behavior as follows. $k_2/k_1$ is the ratio of the rates with which cavity-2 emits a photon compared to cavity-1. As $k_2$ increases relative to $k_1$, cavity-2 emits photons much more frequently than cavity-1, and thus there is less likelihood that those photons will reach the input ports of the beam splitter at the same time. This causes the probability of coincidence to drop to 50\% (a classically expected result) as $k_2/\kappa_1\gg1$.

\begin{figure*}
\centering
    \includegraphics[width=2.5in, height=1.7in]{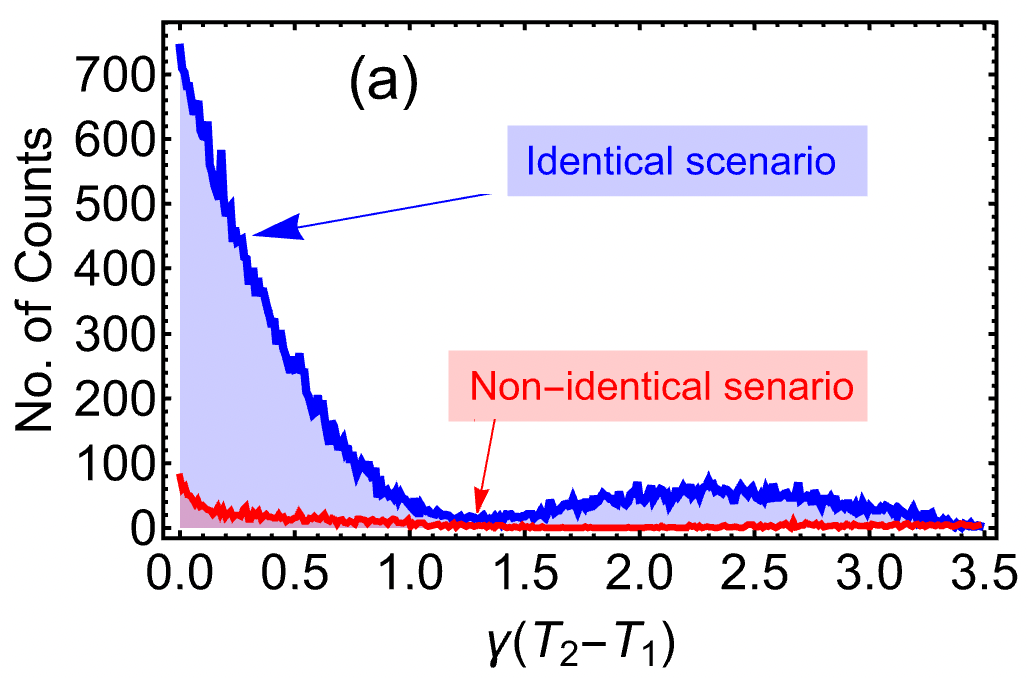}
    \includegraphics[width=2.5in, height=1.68in]{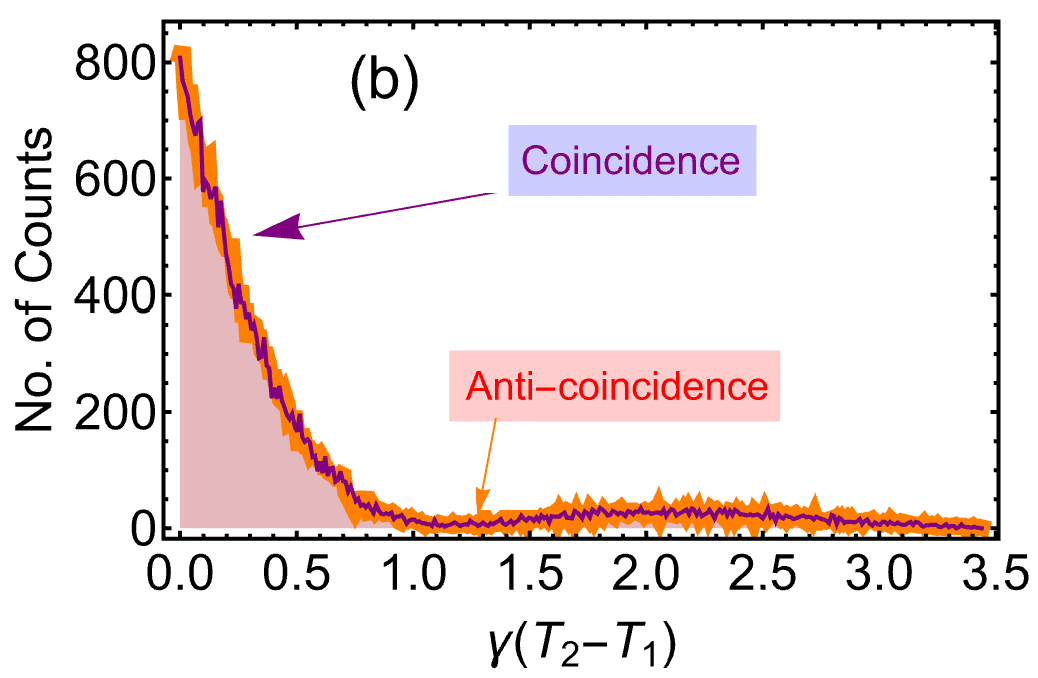}
\captionsetup{
  format=plain,
  margin=1em,
 justification=raggedright,
 singlelinecheck=false
}
\caption{(Color online) (a) Coincidence counts as a function of the time delay between the two clicks recorded at the output photodetector (first/second click time is referred to as $T_1/T_2$). Here have considered an identical scenario (blue curve) where we have set $\kappa_1 = \kappa_2 = g_1 = g_2 = \gamma_1$=$\gamma_2=\gamma$ and a non-identical scenario (red curve) where $\kappa_1 = 0.04\gamma$, $\kappa_2 = 25\gamma$ and the rest of the parameters are the same as in blue curve. (b) The number of coincidence counts (purple curve) and anti-coincidence counts (orange curve) as a function of detection delay time generated by two independent atom-cavity sources. For this plot, we have chosen the same set of parameters as used for the identical scenario in part (a) of the plot.}\label{Fig9}
\end{figure*}

As the final set of results, we focus on the delay time between the detection of the two photons at the output detectors. It is known that when the standard HOME conditions are met, the two photons emerging at the output detectors tend to appear close in the time domain, a phenomenon referred to as the bunching of photons in the quantum optics literature \cite{paul1982photon, bouchard2020two, mirza2015nonlinear}. Here, we study the same phenomenon and report our results in Fig.~\ref{Fig9}. We present two types of plot here, while indirectly controlling the time when the photons appear at the input ports of the beam splitter by the cavity leakage rates $\kappa_1$ and $\kappa_2$. First, in Fig.~\ref{Fig9}(a), we plot the number of coincidence counts in the identical scenario (blue curve), that is, when both atom-cavity sources are identical (in particular, with $\kappa_1=\kappa_2$). For comparison, we also plot the number of coincidence counts in the nonidentical scenario on the same scale, with all other parameters being the same for both atom-cavity subsystems except $\kappa_1=0.04\gamma$ and $\kappa_2=25\gamma$. 

Using the same model for fixed $\kappa_n$ and variable $t$, we found that (as expected) the highest number of coincidences occurred for zero $ t_2-t_1$ and decreased over time. Note that this drop-off occurs faster when $\kappa_2 \neq \kappa_1$, suggesting that for coupled \textit{identical} cavities the Hong-Ou-Mandel effect is stronger than for coupled distinguishable cavities. To verify the photon-photon interference effect, we present our results for the uncoupled or independent cavity QED case in Fig.~\ref {Fig9} (b). Here, we ran our Monte Carlo simulation for a single cQED single-photon source arranged against a beam splitter twice and analyzed the union of these results. As expected, when cavities are entirely independent, the probability of receiving a coincidence versus anticoincidence event at a given point in time is nearly equal, confirming the lack of any interference between the photons.

%%===================================================%%
%%         Sec.VI: Summary and Conclusions           %%
%%===================================================%%
\section{\label{sec:IV} Summary and Conclusions}
In this paper, we studied the HOME with single-photon sources based on initially excited two-level atoms and cavity QED systems. Our theoretical tool (for both analytic and numerical calculations) was the quantum jump/trajectory approach. Our analytic results confirmed the presence of HOME for both types of single-photon source, with the possibility of generating bipartite-entangled Bell states for the excited-atom problem after the detection of the first quantum jump. Our numerical results for cavity QED sources indicate the presence of Rabi oscillations in the time-dependent average photon and excitation number plots. Therein, the coincidence counts showed the lack of pure quantum interference effects when the two atom-cavity systems have different leakage rates. Finally, our results on detection time delays indicated a bunching of photons for zero delays in the case where both atom-cavity systems were identical and coupled. The uncoupled or independent atom-cavity situation led to equal coincidence and anticoincidence counts, thereby confirming the absence of any two-photon interference effects. These interesting results have the potential to be beneficial in the areas of linear optics, quantum computing, and quantum information science protocols that rely on the indistinguishability of photons.

%%===================================================%%
%%          Article Acknowledgements                  %%
%%===================================================%%
\section*{Acknowledgments}
IMM acknowledges financial support from the NSF LEAPS-MPS 2212860 Grant and start-up funding from Miami University of Ohio, USA.

\bibliographystyle{ieeetr}
\bibliography{paper}
\end{document}